\documentclass[conference]{IEEEtran}
\IEEEoverridecommandlockouts
\usepackage{cite}
\usepackage{amsmath,amssymb,amsfonts}
\usepackage{algorithmic}
\usepackage{graphicx}
\usepackage{textcomp}
\usepackage{booktabs}       % professional-quality tables
\newcommand{\ra}[1]{\renewcommand{\arraystretch}{#1}} % More space between rows
\usepackage{xcolor}
\usepackage{xspace}
\usepackage{url}
\newcommand{\PaperAcronym}{RTClean\xspace}
\newcommand{\quotes}[1]{``#1''}
\usepackage{listings}
\usepackage[color=gray!17]{todonotes}
\let\xtodo\todo
\renewcommand{\todo}[1]{\xtodo[inline,color=black!5]{#1}}

\usepackage{cleveref}
\def\BibTeX{{\rm B\kern-.05em{\sc i\kern-.025em b}\kern-.08em
    T\kern-.1667em\lower.7ex\hbox{E}\kern-.125emX}}
\begin{document}

\title{\PaperAcronym: Context-aware Tabular\\ Data Cleaning using Real-time OFDs
%\thanks{}
}

\author{\IEEEauthorblockN{Daniel Del Gaudio}
\IEEEauthorblockA{\textit{University of Stuttgart}\\
Stuttgart, Germany \\
Daniel.Del-Gaudio@ipvs.uni-stuttgart.de}
\and
\IEEEauthorblockN{Tim Schubert}
\IEEEauthorblockA{\textit{University of Stuttgart}\\
Stuttgart, Germany \\
st148736@stud.uni-stuttgart.de}
\and
\IEEEauthorblockN{Mohamed Abdelaal}
\IEEEauthorblockA{\textit{Software AG} \\
Darmstadt, Germany \\
Mohamed.Abdelaal@softwareag.com}
}

\maketitle

\begin{abstract}
Nowadays, machine learning plays a key role in developing plenty of applications, e.g., smart homes, smart medical assistance, and autonomous driving. A major challenge of these applications is preserving high quality of the training and the serving data. Nevertheless, existing data cleaning methods cannot exploit context information. Thus, they usually fail to track shifts in the data distributions or the associated error profiles. To overcome these limitations, we introduce, in this paper, a novel method for automated tabular data cleaning powered by dynamic functional dependency rules extracted from a live context model. As a proof of concept, we create a smart home use case to collect data while preserving the context information. Using two different data sets, our evaluations show that the proposed cleaning method outperforms a set of baseline methods in terms of the detection and repair accuracy.
\end{abstract}


% Insert sections
%=====================
\section{Introduction}\label{sec:introduction}
%=====================
%
% introducing machine learning and data quality
\textit{Machine learning and data quality:} Over the last couple of decades, the technological advances in storage and processing power have broadly enabled numerous innovative products and services based on machine learning (ML), e.g., Facebook's facial recognition program, Google's translation and speech recognition, Netflix’s recommendation engines, Uber's dynamic pricing systems, and Tesla's self-driving cars. In such applications, machine learning is typically employed to derive insightful information from (large volumes of) data by leveraging algorithms to identify complex patterns and learn in an iterative process. One of the essential factors in training an effective machine learning model is providing high quality data. In fact, poor data quality can lead to incorrect business intelligence decisions, worse data analysis, and a multitude of errors. According to Gartner research, organizations believe poor data quality to be responsible for an average of \$15 million per year in losses \cite{gartner}.

% introducing data errors and problems of static data cleaning
\textit{Challenges:} Several error types, e.g., numerical outliers, null values, rules/constraints violation, typos, duplicates, and inconsistencies, may co-exist in real-world tabular data. Such error types typically originate owing to improper join operations, noisy communication channels, inaccurate and incomplete manual data entry, etc. To combat data quality problems, several error detection methods have been introduced \cite{raha,holoclean,metadrive2018} to automatically identify erroneous data instances, before either removing or imputing them. These methods usually leverage a set of \textit{static} cleaning signals, e.g., business rules, data constraints, or metadata, to detect different error types. Nevertheless, such cleaning signals usually lack information about the context of data collection, which can play an important role while cleaning the data.  

%Specifically, these signals are typically collected in an offline phase, e.g., during the development phase of ML models. Accordingly, these signals lack information about the dynamic environments in which the ML models, trained on the cleaned data, will be used. In this case, possible changes to any of these cleaning signals, e.g., during the deployment phase, may drastically deteriorate the accuracy of the detection decisions made by such methods.

\textit{Motivating Scenario:} To further explain the challenges of dynamic environments while cleaning tabular data, we created a \textit{Smart Home} use case. From this use case, we collected a real-world data set (cf. evaluations in Section~\ref{sec:evaluation}). Our environment consists of four temperature sensors: two are placed in the same room, one in another room and one outside the Smart Home. Each sensor records the temperature value every hour. The collected data has been used to predict the energy consumption of our environment using ML models. Since the sensors tend to produce erroneous values from time to time, the predictions of the trained ML models are broadly not accurate. A manual detection of errors in the collected data is an overwhelming and time-consuming task, yet, context knowledge about the environment that the sensors are placed in can be helpful. For instance, two temperature sensors in the same room, measuring different values at the same time, can be an indication of an error. In fact, existing data cleaning methods cannot exploit such context knowledge to enhance their performance. 

% introducing our poposed approach and motivate for the need to dynamic rules in the realm of data cleaning.
\textit{Context-aware cleaning:} To overcome the limitations of existing data cleaning methods, we introduce, \textit{\PaperAcronym}\footnote{RTClean: https://github.com/delgaudl/RTClean}, a novel data cleaning method which considers the context of data collection and the dynamic nature of the deployment environments. To this end, \PaperAcronym leverages a context model of the intended application to capture a set of \textit{Ontological Functional Dependencies} (OFDs). Such OFDs describe semantic attribute relationships such as synonyms and is-a hierarchies defined by an ontology. It is important to highlight that several previous works exploited OFDs for data cleaning \cite{ofd2020,ofd2022} (cf. Section~\ref{sec:related_work} for more details). However, \PaperAcronym differentiates itself from these works in its ability to track real-time changes in the environment, e.g., relocating the sensors in the Smart Home scenario, thanks to our \textit{live} context model. Specifically, our context model enriches the static information, embedded in an ontology, with live monitoring and sensor data. Since the environment is mostly dynamic in terms of the available devices and sensors, our context model can be updated constantly. Note that we do not aim to use OFDs on the sensor data itself but on the context of the environment in which the data is generated.

% summarize the contributions
\textit{Contributions:} To sum up, the paper provides the following contributions: (1) We introduce a novel three-steps error detection method, including context modeling, feature generation, and error detection. Based on our live context model, the error detection method can be updated with fresh OFDs. (2)~We introduce a Smart Home use case to collect an IoT data set, while preserving the context information. (3) We evaluate the performance of \PaperAcronym using two real-world data sets, including our Smart Home data set and the Hospital data set from the US health service\footnote{Hospital data set: https://gitlab.com/hatjog/holoclean/-/tree/master/testdata}, to ensure the generality of the proposed method. In both data sets, the results show that \PaperAcronym outperforms a set of baseline methods. To the best of our knowledge, \PaperAcronym is the first data cleaning method which can be updated in real-time via considering both static and dynamic information about the environments.
% 
% define the structure of the paper
%\textit{Paper's structure:} The remainder of the paper is structured as follows. Section~\ref{sec:overview} provides an overview of the proposed solution, together with highlighting our assumptions. Section~\ref{sec:ontology} introduces our ontology-based context model, before we elaborate on our proposed context-aware data cleaning method in Section~\ref{sec:contextual_cleaning}. Section~\ref{sec:prototype} presents the implementation details of our proposed method. Finally, Section~\ref{sec:evaluation} summarizes the results of our evaluations, before Section~\ref{sec:conclusion} draws a conclusion of the obtained results with an outlook on future work. 
%=============================
\section{System Model and Architecture}\label{sec:overview}
\begin{figure}[t]
	\centerline{\includegraphics[width=0.5\textwidth]{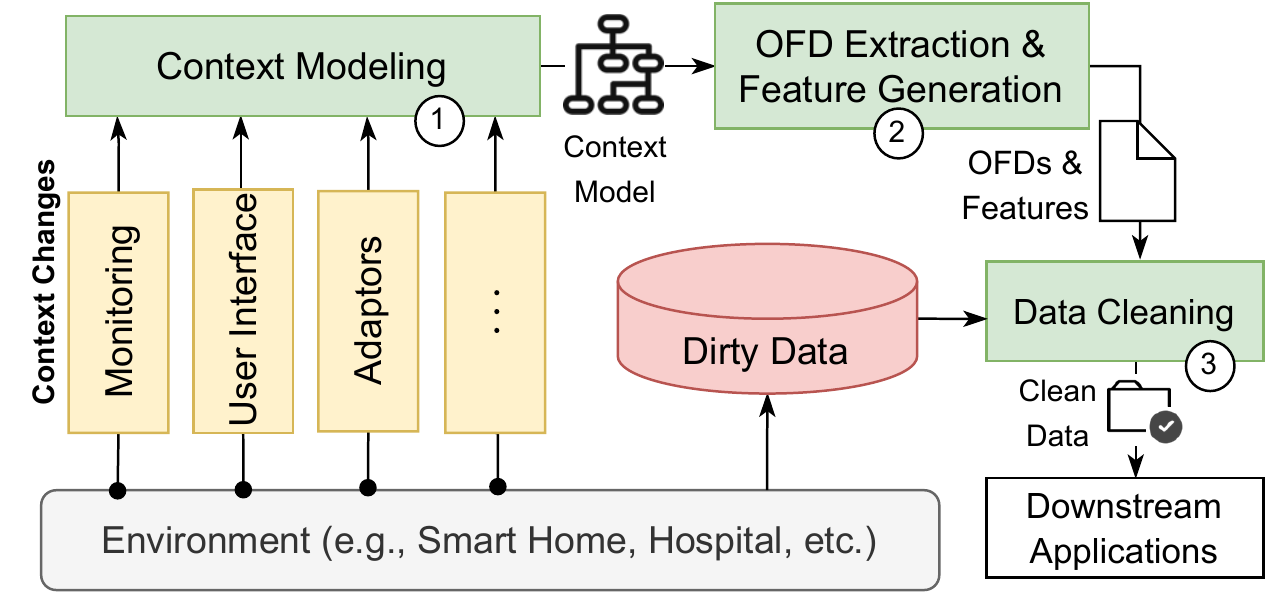}}
	\caption{Overview of the \PaperAcronym approach.}
	\label{fig:overview}
\end{figure}

%=============================
%
Our main objective is to enrich existing data cleaning methods with features extracted from an ontology-based context model. \PaperAcronym is divided into three steps, which also reflect the architecture of our system, since we propose a separate component for each step. Thus, each component can be further involved into different processes, e.g., using the context model for system engineering. Figure~\ref{fig:overview} provides an overview over all components of \PaperAcronym with their respective inputs. As the figure shows, \PaperAcronym encompasses three main steps, which are defined as follows.

\textbf{1) Context modeling:}
The ontology-based context model must be manually/automatically generated in an offline phase (It has been generated automatically for our Smart Home use case). The concept to monitor an environment, detect changes and update the context model constantly, using different adapters, relies on our previous work~\cite{9767267} and is therefore beyond the scope of this concept. However, it is important to notice, that in our concept, context changes can originate from multiple different sources, e.g., by the interaction of human users via a user interface.

\textbf{2) OFD extraction and feature generation:}
In the next step, a set of different OFDs are extracted from the ontology. To this end, \PaperAcronym relies on the OFD discovery algorithm proposed in \cite{10.1145/3132847.3132879}. From the extracted OFDs, a set of binary features are generated that can be used for data cleaning.

\textbf{3) Data cleaning:}
Finally, the dirty data can be cleaned through exploiting the binary features generated from the OFDs dependencies. This step outputs a cleaned data set and a set of detected errors. Afterward, the clean data can be employed in different downstream applications, e.g., ML model building, visualizations, business intelligence, etc.

When the whole system is initiated, changes in the environment are constantly registered using different adapters. Thus, if we have a stream of data originating from the environment, we can adapt the data cleaning step in real-time to the current context. In this paper, we assume that \PaperAcronym receives a data set, that contains errors, and are able to create an ontology-based context model that describes the structure of the data or of the environment which the data is extracted from. For a better understanding of the remainder of this paper, Section~\ref{sec:ontology} gives an overview of the context model described in~\cite{9767267} together with the extensions we made to use the context model for data cleaning.
%=====================================
\section{Ontology-based context model}\label{sec:ontology}
%=====================================
%
\PaperAcronym aims to use knowledge about the context of data to improve data cleaning methods. To this end, \PaperAcronym leverages an ontology-based context model to consolidate the context knowledge, which is described in this section. OFDs can then be used to generate additional features from the model. For the context model, we use the live context model approach that we proposed in our previous work~\cite{9767267}, which builds on the ontologies IoT-Lite~\cite{7816831} and SSN~\cite{COMPTON201225}. The use of ontologies enables the extensibility of the context model to different data and use cases. The approach aims to adapt the ontology to the current state of the system. This adaptation is achieved by using different adapters that periodically retrieve changes in the environment, e.g., new devices, and update the context model accordingly. Such fresh context information enables us to adapt the data cleaning automatically to changes of the environment in real-time. We extended the ontology proposed in~\cite{9767267} for the purpose of data cleaning. 
\begin{figure*}
\centerline{\includegraphics[width=1\textwidth]{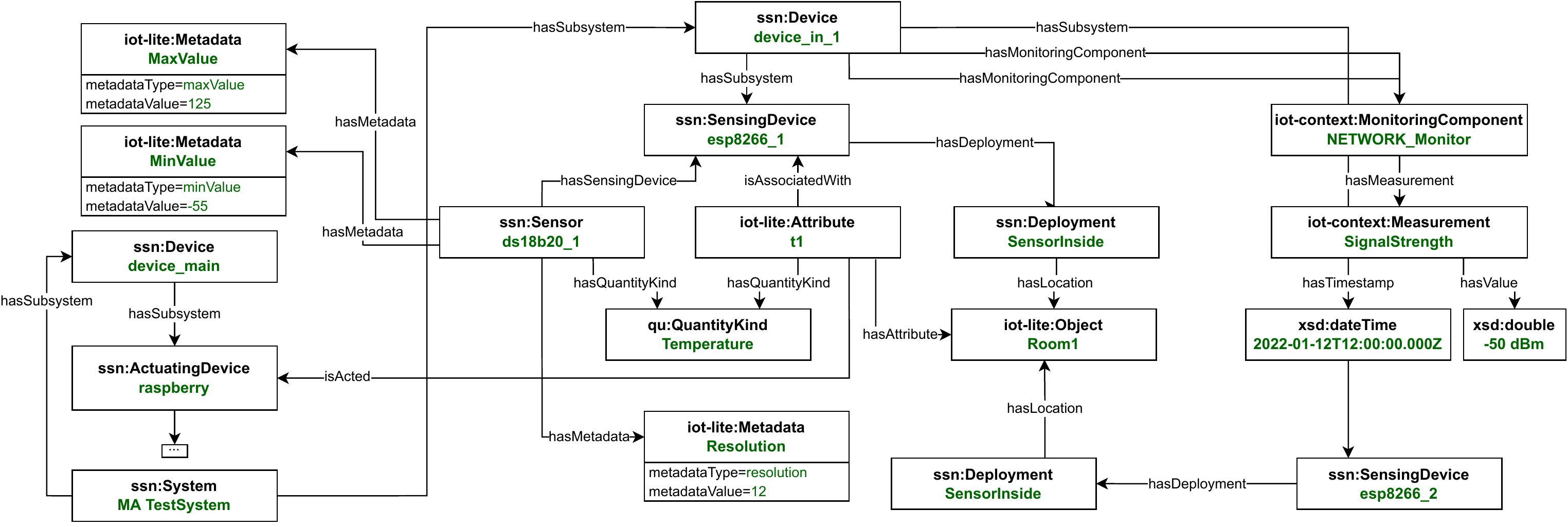}}
\caption{Extract of the IoT context model ontology.}
\label{fig:ontology}
\end{figure*}

\begin{figure}
	\centering
	\includegraphics[width=\linewidth]{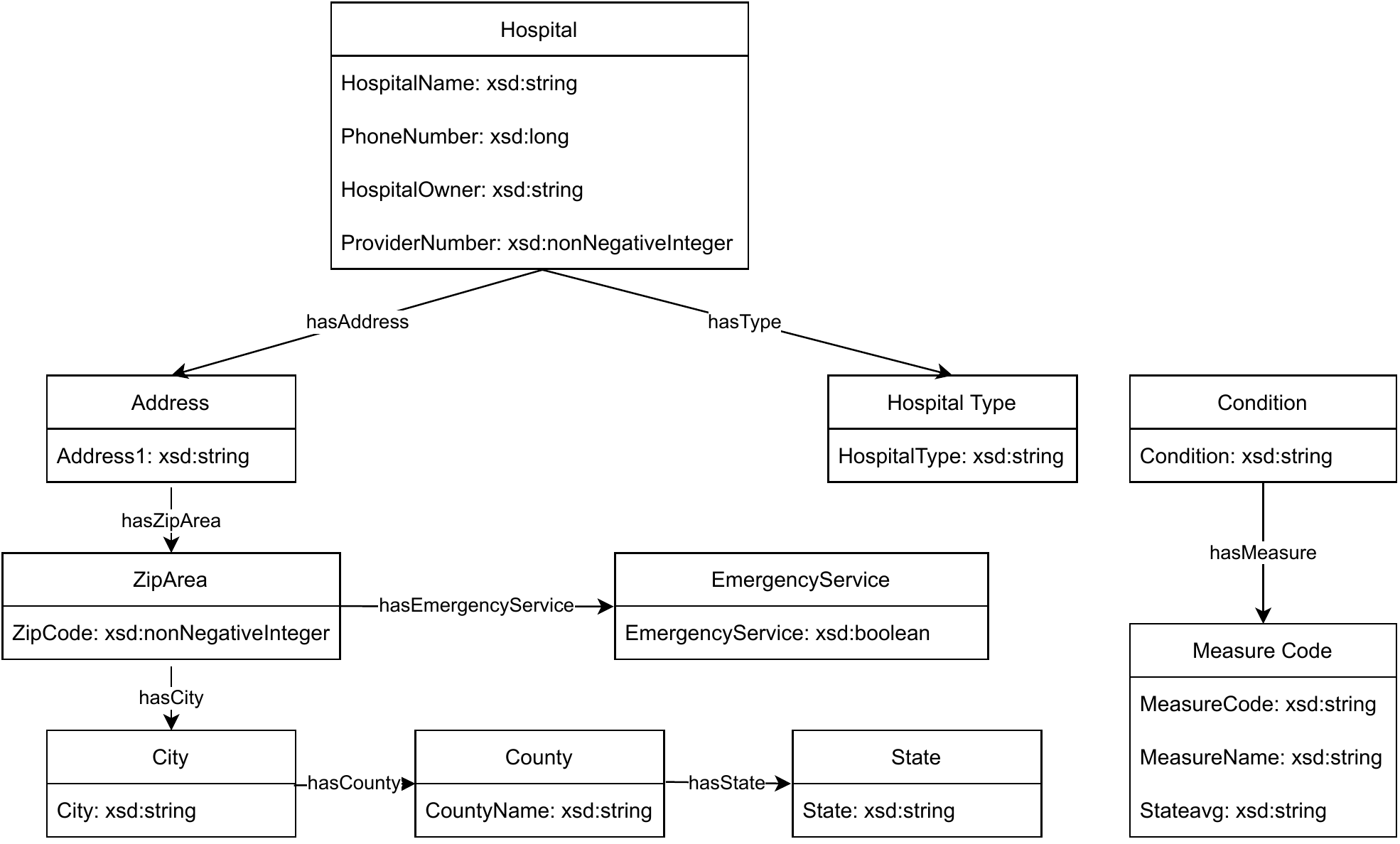}
	\caption{Ontology to represent the structure of the Hospital data set.}
	\label{fig:hospital-ontology}
\end{figure}

Figure~\ref{fig:ontology} shows an extract of the context model of the IoT scenario, described in Section~\ref{sec:introduction}. The \textit{System} is the root node and indicates one environment. While, the \textit{Devices} node indicates physical devices that are currently present in the environment. As the figure depicts, the \textit{Devices} are linked with the \textit{Sensors}, which indicate currently available sensors, by the use of \textit{Sensing Devices}. The \textit{Sensors} can be attached with \textit{Metadata}, e.g., to indicate maximum, minimum values and the resolution of the sensor. Each \textit{Sensing Device} can be associated with a location, for which we use the \textit{Objects}, via a \textit{Deployment} node. It is important to note, that the structure of the context model highly depends on the structure of the data, the use case and the environment from which the data is originated. Therefore, we utilize another data set for the evaluation with an according context model. \Cref{fig:hospital-ontology} shows the ontology of the Hospital data set. 

%====================================
\section{Context-aware Data Cleaning}\label{sec:contextual_cleaning}
%====================================
%
In this section, we elaborate on the context-aware data cleaning method. First, we give a definition of the various OFDs dependencies which can be extracted from our live context model. Second, we explain how \PaperAcronym exploits these automatically-generated OFDs dependencies to detect and repair erroneous data instances.

%-------------------------------------------------
\subsection{OFD Extraction from Ontologies}
%-------------------------------------------------
%
In general, OFDs are a special kind of \textit{Functional Dependencies}~(FDs), derived from an ontology. The different types of OFDs that \PaperAcronym focuses on are explained in this section. Considering a data set $D$ with a relational schema $R$. An FD $X \to Y$ is a constraint which uniquely determines the relation of an attribute $Y \in R$ to a set of attributes $X \subseteq R$~\cite{statperspectiveFD}. They mostly describe relationships based on syntactic equality and can be used, e.g., in data integration and data cleaning \cite{discoveryofd}. In this context, the FD dependencies are used to specify data quality requirements. An instance $I$ of $R$ satisfies the $FD$ $X \to Y$ if for every pair of tuples $t_1, t_2 \in I$ if $t_1[X]=t_2[X]$, then $t_1[Y]=t_2[Y]$. For example, consider the following two attributes: \textit{City} and \textit{ZipCode}. If the tuples $t_1[city] = t_2[city] =$ \quotes{Berlin}, then the corresponding tuples $t_1[ZipCode], t_2[ZipCode]$ should be equal to 10115.

To derive FDs from data observations, identification of the attribute order which defines the directionality is needed. In order to reduce the exponential computational cost, the existing methods rely on pruning to efficiently search over the lattice of attribute combinations~\cite{statperspectiveFD}. From our live context model, \PaperAcronym can extract different dependencies, including \textit{structure-based}, \textit{time-based} and \textit{value-based} dependencies. The time-based and value-based dependencies differentiate our context model from other relevant works, e.g., \cite{ofd2020} and \cite{ofd2022}. The OFDs are extracted by querying the context model for the different types of OFDs. Below, we describe each type of these dependencies. 

%------------------------------
\subsubsection{Structure-based}
%------------------------------
%
The extraction of structure-based dependency usually does not consider the actual values of the data, but solely the structure of the ontology. In this context, we consider three different kinds of structure-based dependencies: \emph{denial dependencies}, \emph{matching dependencies}, and \emph{device link dependencies}. The denial dependencies contain, e.g., functional and conditional dependencies, which indicate whether an instance in the data set is erroneous, if the dependency is satisfied. A denial dependency $D$ over a relation $R$ is defined as $A \to B$, where $A$ and $B$ are single attributes in $R$ and $A \neq B$. An instance $I$ satisfies $D$ if for every pair of tuple $t_1, t_2 \in I$ with $t_1[A] = t_2[A]$, $t_1[B] \neq t_2[B]$.

On the other hand, the matching dependencies generalize FDs by requiring pairs of tuples to be similar w.r.t. a certain similarity metric in the left- and right-hand side column values in lieu of being strictly equal. A matching dependency $MD$ over a relation $R$ is represented as $A_{x\%} \to B_{{x\%}}$, where $A$ and $B$ are \textit{single} attributes in $R$ and $A \neq B$. The subscript $x\%$ denotes the level of similarity between $A$ and $B$. An instance $I$ satisfies $MD$ if for every pair of tuple $t_1, t_2 \in I$ with $t_1[A] = t_2[A]$, similarity between $t_1[B], t_2[B] \geq x\%$. For the device link dependency, it indicates whether a sensor is linked to a specific device. A device link dependency $L$ can be represented as a function $\Psi: X \to Y$, where $X$ denotes the set of sensors available in the IoT environment and $Y$ denotes devices. A dependency $\Psi(A) = B$ is satisfied, if the sensor $A$ is physically connected to the device $B$. This relation means that the sensor itself belongs to this device and can only be read out from it specifically.

%-------------------------
\subsubsection{Time-based}
%-------------------------
%
\emph{Temporal dependencies} are used to detect errors in the order of the data. To detect errors concerning this dependency, the data must be equipped with a timestamp. Specifically, temporal dependencies $T$ define a relation between two devices, $A$ and $B$. If the dependency $A \to B$ holds, the data, e.g., measurements, will only be sent from device $A$ to $B$. Let the timestamp $t_A$ denotes the time when a message $m$ is created or processed on the device $A$. Since the transmission time from device $A$ to device $B$ is greater than zero, the timestamp $t_A$ should be smaller than the timestamp $t_B$ (timestamp at which the message is processed on the device $B$). In this case, the device $A$ can be called a temporal predecessor of $B$.

%--------------------------
\subsubsection{Value-based}
%--------------------------
%
We consider three different kinds of value-based dependencies: \emph{locality dependencies}, \emph{monitoring dependencies}, and \emph{capability dependencies}. A locality dependency $L$ is defined as a function $\Gamma: Device \to Locality$. Let $A$ be a sensing device and $B$ a locality, e.g., a room. If the device $A$ is placed at the location of $B$, then the locality dependency $A \to B $ holds and $\Gamma(A)=B$. This means that measurements taken from the sensing device $A$ will always capture the environment at the location $B$. A monitoring dependency $M$ describes a device $A$, which is monitored by a monitoring component $B$. For instance, health indicators, e.g., the CPU load, can be stored as measurements in the context model. Since those measurements are being added live while operating, monitoring dependencies can be used in real-time or on an existing data set with timestamps. 

A capability dependency $C$ describes a set of capabilities $B$ assigned to a sensor $A$. One capability is represented as a metadata object for a specific sensor. The metadata object consists of a type, e.g., \textit{resolution} or \textit{minimal measurable value}, and the corresponding value. Since metadata objects store the capabilities of a sensor, they can be used as a filter for the measured values. Specifically, the values which are not plausible, regarding the \textit{sensor's abilities}, can then directly be marked as erroneous. For example, if a measurement is lower than the minimal measurable value of a sensor, this measurement is considered as erroneous. A capability can also define borders of the \textit{measured units}. For example, temperatures which are lower than the absolute zero are then automatically labeled as an error.

%------------------------
\subsection{Data Cleaning}
%------------------------
%
For data cleaning, \PaperAcronym{} is built on top of a state-of-the-art data cleaning framework, referred to as  \textit{HoloClean}~\cite{holoclean}. However, It is important to mention that \PaperAcronym{} can be integrated with any other data cleaning method. We selected HoloClean since its mechanism supports manually-crafted denial and matching dependencies. HoloClean includes several methods for error detection, e.g., null detector, violation detector (using manually-crafted dependencies), and error loader detector\footnote{HoloClean: https://github.com/HoloClean/holoclean}. To exploit the OFDs, extracted from our live context model, in the process of data cleaning, we added an OFD violation detector to HoloClean. Such an OFD violation detector is able to extract the OFDs and then evaluates them on the data set. Specifically, the OFD detector creates a Boolean feature for each OFD, stating whether the OFD is fulfilled for each row in the data. Such binary features serve as the detected errors by the OFD detector. These detected errors are then further used by HoloClean in its normal pipeline.
%\input{sections/prototype}
%===================
\section{Performance Evaluation}\label{sec:evaluation}
%===================
%
In this section, we assess the effectiveness of \PaperAcronym relative to a set of baselines. First, we introduce the experimental setup, before discussing the obtained results
%
%==========================
\subsection{Experimental Setup}\label{sec:prototype}
%==========================
%
To evaluate \PaperAcronym, we utilized two real-world data sets, including the Hospital data set and the IoT data set. Table~\ref{tab:datasets} provides information about the data sets and examples of the extracted OFDs. We utilized an \textit{Error Generator}\footnote{Error Generator: https://github.com/BigDaMa/error-generator} to inject realistic errors in the data sets. The errors comprise \textit{typos}, \textit{value errors}, and \textit{null values}. Regarding the IoT data set, we injected 5\% errors of each category, resulting in 149 erroneous instances. Furthermore, we injected outliers into the numerical values of the IoT data set to compare our approach with a typical outlier detection method. We created two different data sets with numerical values and errors: one with random errors in the original value range, and one with random errors in the doubled range of the original values.
\begin{table}
\centering
\ra{1.2}
\caption{Examples of the extracted OFDs}
\label{tab:datasets}
\resizebox{\columnwidth}{!}{%
\begin{tabular}{@{}lll@{}}
\toprule
                      & \textbf{IoT}                                                                                   & \textbf{Hospital} \\ \midrule
Size                  & (1000,7)                                                                                       & (1000,20)         \\
\# Manual Dependencies & ---                                                                                            & 15                \\
\# Auto-extracted OFDs & 21                                                                                             & 25                \\
OFD: Denial &
  \begin{tabular}[c]{@{}l@{}}System $\to$ Device,\\ Device $\to$ SensingDevice\end{tabular} &
  \begin{tabular}[c]{@{}l@{}}PhoneNumber $\to$ Address1,\\ ZipCode $\to$ City\end{tabular} \\
OFD: Matching &
  --- &
  \begin{tabular}[c]{@{}l@{}}ProviderNumber$_{75\%}$ $\to$ PhoneNumber$_{75\%}$,\\ Stateavg$_{75\%}$ $\to$ MeasureCode$_{75\%}$\end{tabular} \\
OFD: Device-Link      & ds18b20\_1 $\to$ device\_in\_1                                                                 & NA               \\
OFD: Capability       & \begin{tabular}[c]{@{}l@{}}ds18b20\_1 $\to$ MaxValue,\\ ds18b20\_1 $\to$ MinValue\end{tabular} & NA               \\
OFD: Locality         & \begin{tabular}[c]{@{}l@{}}ds18b20\_1 $\to$ Room1,\\ ds18b20\_2 $\to$ Room1\end{tabular}       & NA               \\
OFD: Temporal         & device\_in\_1 $\to$ device\_main                                                               & NA        \\ \bottomrule      
\end{tabular}%
}
\end{table}

It is important to mention that the original implementation of HoloClean supports only denial dependencies and, with a workaround, matching dependencies. We extended this implementation to exploit the extracted OFDs. As baseline methods, we utilized \textit{Raha}~\cite{raha}, an ML-based error detection method, and \textit{dBoost}\footnote{dBoost: https://github.com/cpitclaudel/dBoost} which is an outlier detection method for numerical values. For the evaluation, we carried out data cleaning in three different ways: (i)~without any dependencies (using only the null detector already exists in HoloClean), (ii)~with manually-crafted dependencies, and (iii)~with automatically-extracted dependencies. We store the context model in an Apache Jena\footnote{Apache Jena: https://jena.apache.org/index.html} triple store database and use the Fuseki\footnote{Apache Fuseki: https://jena.apache.org/documentation/fuseki2/index.html} SPARQL server to extract the dependencies. The evaluation metrics comprise the detection precision, recall and F1-score. Furthermore, we employ the repair recall, defined as the fraction of correct repairs over the total number of correctly-detected erroneous cells~\cite{holoclean}, and the repair F1-score, defined as the harmonic mean of precision and repair recall. 
\subsection{Results}
%===================
%
%------------------------------
\subsubsection{Hospital Data Set}\label{sec:hosp_dataset}
\begin{figure}
	\centerline{\includegraphics[width=1\columnwidth]{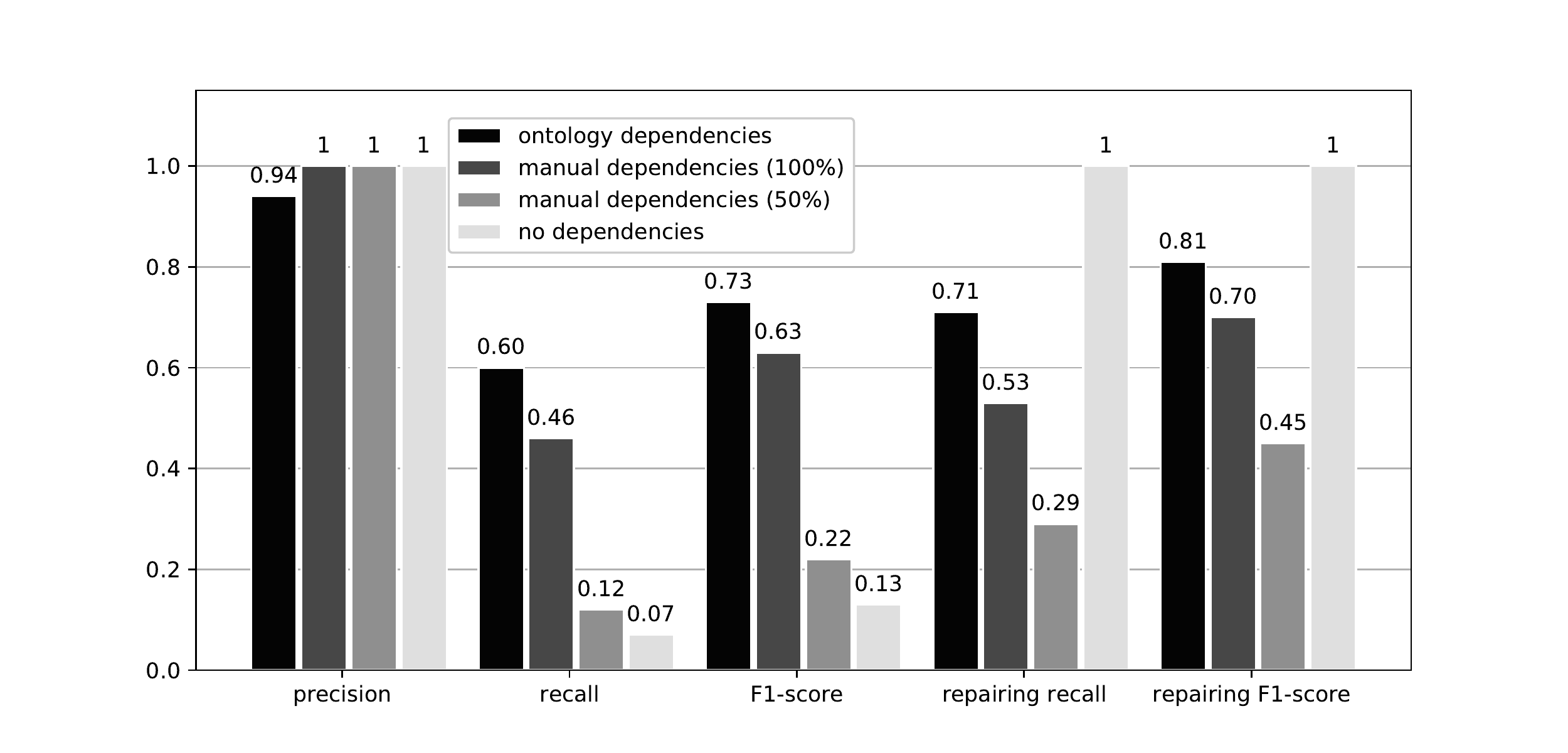}}
	\caption{Results for the Hospital data set in comparison with HoloClean using 50\% manual dependencies, 100\% manual dependencies and no dependencies.}
	\label{fig:results_hospital}
\end{figure}

%------------------------------
%
The Hospital data set contains names of hospitals with their addresses, their types, and other categorical attributes, as it can be seen in Figure~\ref{fig:hospital-ontology}. The data set comes with a set of manually-created dependencies, and it includes a set of erroneous instances. Figure~\ref{fig:results_hospital} shows the results for the Hospital data set. For the comparison, we executed the original implementation of HoloClean on the Hospital data set with no dependencies, 50\% of the manual dependencies, 100\% of the manual dependencies, and the dependencies automatically extracted from the context model. 

As Figure~\ref{fig:results_hospital} depicts, the detection and repair accuracies are improved when increasing the number of dependencies from 50\% to 100\%. Such a result emphasizes the general impact of using dependencies for data cleaning. Aside from the number of dependencies, HoloClean with no dependencies achieves high precision. However, it detects only 36 errors in total, whereas \PaperAcronym detects 431 errors in total. Similarly, the high precision of HoloClean with no dependencies leads to a high repair recall and repair F1-score, because all 36 errors have been correctly repaired. In the case of using manually-crafted dependencies, HoloClean also achieves high detection precision, since the dependencies have been created specifically for the errors in the data set. However, they achieve low detection recall, since the available dependencies are not sufficient to describe the important relationships in the data set. Conversely, \PaperAcronym yields a significant improvement in recall, F1-score, repair recall and repair F1-score thanks to the automatically-generated OFDs. For instance, \PaperAcronym achieves higher detection F1-score, at least by 14\%, relative to HoloClean with 100\% dependencies.
\subsubsection{IoT Data Set}
\begin{figure}
	\centerline{\includegraphics[width=0.9\columnwidth]{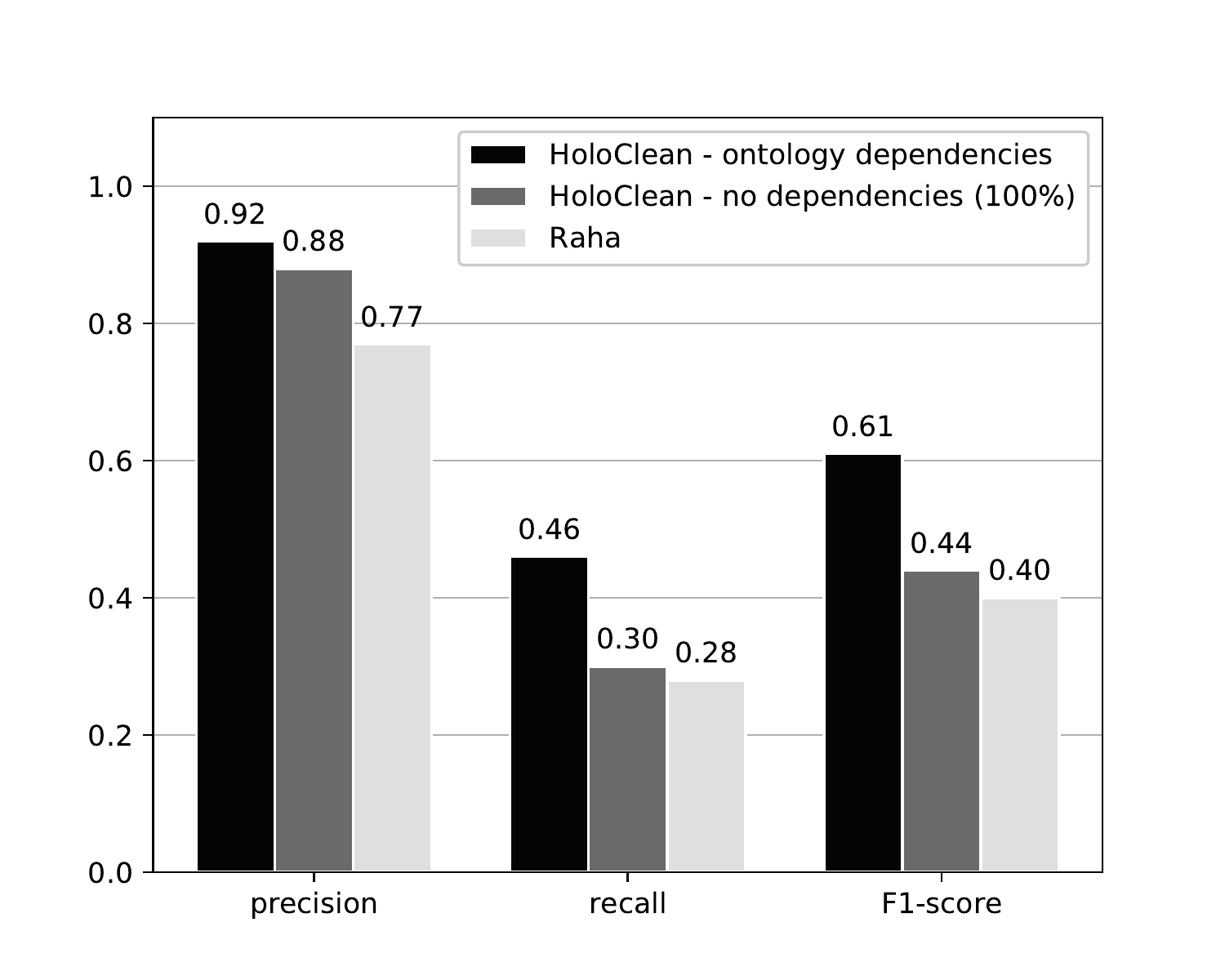}}
	\caption{Results for the whole IoT data set in comparison with \textit{Raha}.}
	\label{fig:results_iot}
\end{figure}

%-------------------------
%
The IoT data set contains temperature read-outs every hour. If more than one value is recorded in this interval, the mean is calculated. The data set is mostly free from errors, but it contains null values and the read-out value of -128 is resulted when a sensor is faulty. These already existing errors are removed manually before the evaluation to have cleaner results for comparison. Figure~\ref{fig:results_iot} shows the results for the whole IoT data set, including the numerical and categorical attributes. We use Raha and HoloClean without dependencies for comparison. In this case, we do not consider the repair recall and the repair F1-score, because our goal for the IoT scenario is to simply detect and erase the erroneous instances. As the figure depicts, \PaperAcronym broadly outperforms the baseline methods, where it achieves higher detection F1-score at least by 28\% and 35\% than Raha and HoloClean without dependencies, respectively.

\begin{figure}
	\centerline{\includegraphics[width=\columnwidth]{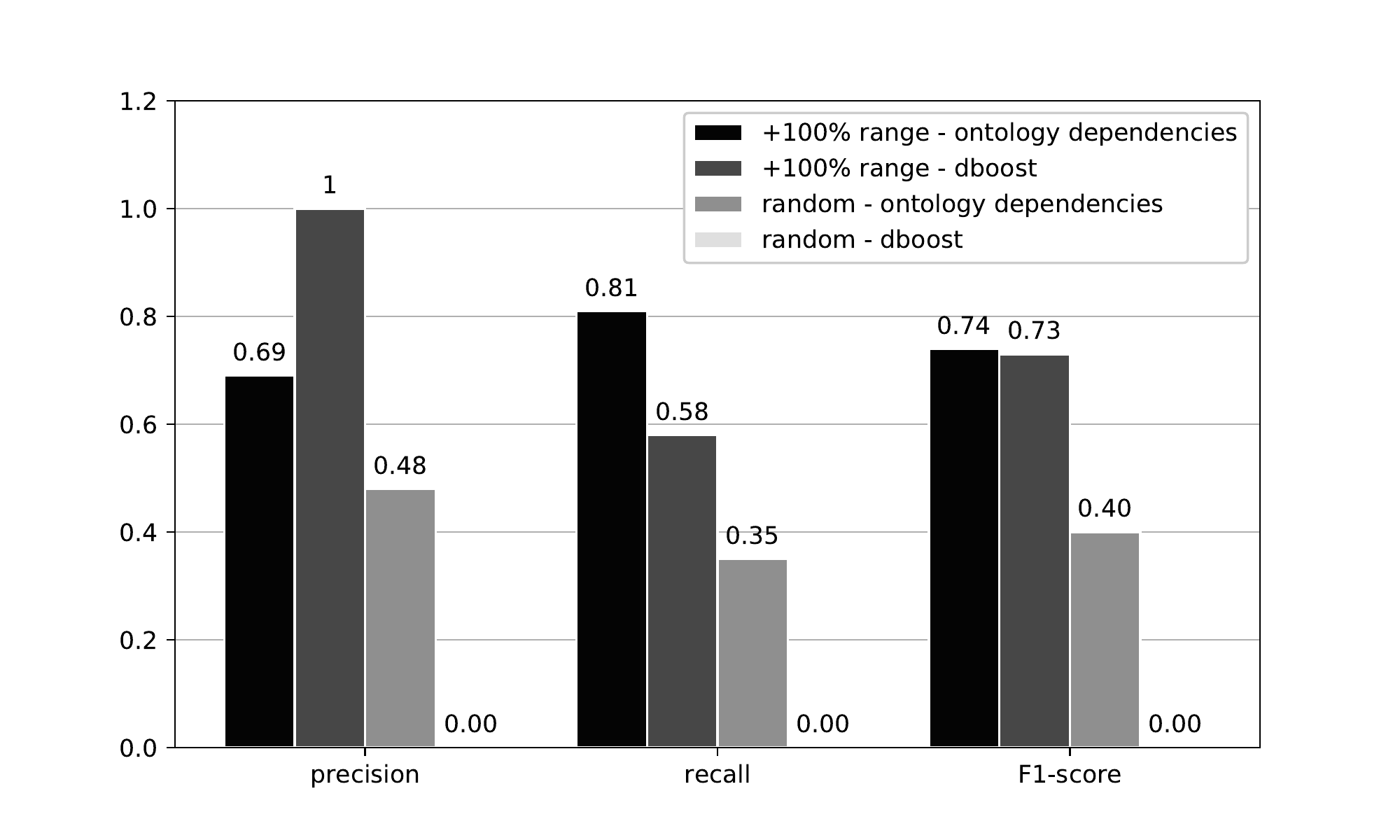}}
	\caption{Results for the numerical values of the IoT data set in comparison with dboost.}
	\label{fig:results_iot_outlier}
\end{figure}

To measure the ability of \PaperAcronym to detect numerical outliers with different ranges, we carried out a comparison between \PaperAcronym and \textit{dBoost}. Figure~\ref{fig:results_iot_outlier} shows the detection accuracy of \PaperAcronym and dBoost while being used for cleaning the IoT data set. In this figure, the label ``random'' indicates the IoT data set with errors in the original range of values. Whereas, the label ``+100\%'' indicates the data set with errors in the doubled range of the original values. As the figure depicts, dBoost performs well in terms of the detection precision on the ``+100\%''-data set. However, it fails to detect erroneous instances when they are in the original range. Conversely, \PaperAcronym can effectively detect errors in the different ranges. For instance, \PaperAcronym achieves higher detection F1-score (at least by 1.3\% and 40\%) than dBoost in the ``+100\%'' range and the original range, respectively.

\section{Related Work}\label{sec:related_work}
%=====================
%
Zheng et al.~\cite{ofd2022,discoverydatacleaning} propose an approach for dependency-based data cleaning with OFDs. They show that real-world data contains domain-specific relationships beyond syntactic equivalence or similarity, e.g., synonyms, which can often be described in ontologies.
Similar to \PaperAcronym, they extract such synonym relationships from ontologies using OFDs.
In contrast to their work, \PaperAcronym also focuses on errors concerning numeric values.
Furthermore, we aim to keep the data cleaning process up-to-date using real-time OFDs. Zhang et al.~\cite{ofd2020} recognize the need for FDs to clean data. Hence, they focus on the extraction of FDs directly from the data, instead of using an ontology. They use structured learning with a probabilistic model to extract FDs from noisy data. \PaperAcronym integrates expert knowledge into the data cleaning by extracting OFDs from the context model. We consider the combination of both approaches as a topic for future research.

Visengeriyeva and Abedjan~\cite{metadrive2018} propose two holistic approaches to combine different error detection methods for heterogeneous data from different sources. They state that, the structural heterogeneity of these sources is the origin of data quality problems, like missing values, duplicates, inconsistent data, and outliers. In this context, metadata is necessary to decide which error detection approach to use for a specific data source. \textit{Multi-column dependencies} are an example of such metadata which they adopted for error detection. \PaperAcronym exploits a live context model to extract multiple OFDs and use them for error detection.

%Rekatsina et al.~\cite{holoclean},
%Raha~\cite{raha}
%\todo{Todo: HoloClean and Raha}
%=============================
\section{Conclusion \& Outlook}\label{sec:conclusion}
%=============================
%
We presented a novel approach, \PaperAcronym, that improves tabular data cleaning using an ontology-based context model and OFDs. Thus, knowledge about the environment the data originates from can be involved in data cleaning. Using our live context model, \PaperAcronym can automatically generate multiple OFDs which represent the current state of the environment. Hence, \PaperAcronym broadly relieves the burden of creating reliable dependencies describing the relationships in the data. The evaluation results show that \PaperAcronym performs better than typical state-of-the-art error detection methods. Moreover, \PaperAcronym proves to be effective in detecting errors on numerical values that are inside the typical range of the data. In the future, we aim to test \PaperAcronym in an end-to-end ML scenario on large data sets and continuous streams of data. Furthermore, we plan to investigate how \PaperAcronym can handle changes of the context model during the data cleaning process.

\section*{Acknowledgment}
This research was funded by German Federal Ministry of Education and Research (BMBF)
through grants 01IS17051 (Software Campus program), 02L19C155 and 01IS21021A (ITEA project number 20219).
%It is also supported (in part) by the BMBF through grants 02L19C155, 01IS21021A (ITEA project number 20219).

\bibliographystyle{ieeetr}
\bibliography{main}

\end{document}